\documentclass{iopjournal}
\usepackage{xcolor}

\usepackage{graphicx}
\expandafter\let\csname equation*\endcsname\relax
\expandafter\let\csname endequation*\endcsname\relax
\usepackage{amsmath,amssymb}
\usepackage{hyperref}
\usepackage{dsfont}

\newcommand{\bc}{\begin{center}}
\newcommand{\ec}{\end{center}}
\newcommand{\be}{\begin{equation}}
\newcommand{\ee}{\end{equation}}
\newcommand{\ba}{\begin{array}}
\newcommand{\ea}{\end{array}}
\newcommand{\beq}{\begin{eqnarray}}
\newcommand{\eeq}{\end{eqnarray}}

\newcommand{\e}[1]{\langle {#1}\rangle}
\newcommand{\abs}[1]{\left| {#1}\right|}

\begin{document}

\title{Random spin-3/2 antiferromagnetic Heisenberg chains revisited}

\author{Yen-Tung Lin$^1$, Pochung Chen$^{1,2,3,*}$ and Yu-Cheng Lin$^{4,2,*}$}

\affil{$^1$ Department of Physics, National Tsing Hua University, Hsinchu 300, Taiwan} 
\affil{$^2$ National Center of Theoretical Sciences, National Taiwan University, Taipei 10617, Taiwan} 
\affil{$^3$ Center for Theory-Computation-Data Science Research, National Tsing Hua University, Hsinchu 30013, Taiwan} 
\affil{$^4$ Graduate Institute of Applied Physics, National Chengchi University, Taipei 11605, Taiwan}

\affil{$^*$Author to whom any correspondence should be addressed.}

\email{pcchen@phys.nthu.edu.tw}
\email{yc.lin@nccu.edu.tw}


\begin{abstract}

We revisit the spin-reduction transition in the random spin-$3/2$
antiferromagnetic Heisenberg chain using the tensor-network strong-disorder
renormalization group (tSDRG). The predicted transition separates two
random-singlet (RS) phases: one formed by effective spin-$1/2$ moments at
weaker disorder and the other by full spin-$3/2$ moments at stronger disorder.
Within the Damle--Huse classification, this transition corresponds to an
emergent permutation-symmetric $\mathcal{P}_4$ multicritical point on a
$\mathcal{P}_2$ critical line associated with the RS phases.  We first derive
the critical exponents governing the disorder-averaged bulk and end-to-end
correlation functions in the $\mathcal{P}_n$ universality classes within the
Damle--Huse framework.  Using these theoretical predictions, together with a
diagnostic constructed from the mean and variance of the energy-gap
distribution, we identify the $\mathcal{P}_4$ transition point and
the reentrant behavior of the RS phase.

\end{abstract}



\section{Introduction}
\label{sec:intro}

Random antiferromagnetic Heisenberg chains provide a setting in which disorder
and quantum fluctuations give rise to phases governed by infinite-randomness fixed
points. A paradigmatic example is the random-singlet (RS) phase of the spin-1/2
chain, in which spins form singlets over arbitrarily long distances and rare
long-range singlets dominate disorder-averaged correlations~\cite{Fisher1994}. 
For higher-spin chains, the interplay between
randomness and internal valence-bond structure allows distinct RS phases
and transitions between them. The random spin-3/2 chain is a prominent
example, exhibiting an infinite-randomness critical point between two types of RS
phases, as predicted by Refael, Kehrein, and Fisher~\cite{Refael2002}.  At strong randomness, the full spin-3/2 moments form
random singlets, giving rise to the $\mathrm{RS}_{3/2}$ phase. At weaker randomness, the
low-energy degrees of freedom are instead effective spin-1/2 moments, which
form an $\mathrm{RS}_{1/2}$ phase. 
For integer-$S$ chains, by contrast, imposed dimerization is required to realize 
an RS phase at weak randomness, since the gapped phases of the corresponding clean, undimerized chains
prevent the system from flowing toward an infinite-randomness fixed point under weak disorder.

A framework developed by Damle and Huse~\cite{DamleHuse2002}, extending Fisher's strong-disorder
renormalization group (SDRG) approach~\cite{Fisher1994}, classifies infinite-randomness fixed
points in random Heisenberg models with spin $S\ge 1/2$ into universality
classes characterized by permutation-symmetric fixed points, denoted by
$\mathcal{P}_n$.
In the domain-wall description of the Damle-Huse framework, 
a spin-$S$ chain supports $2S+1$ valence-bond-solid
(VBS) domains, labeled by $(\sigma,2S-\sigma)$ (or equivalently by type-$\sigma$), with
$\sigma \in \{0,1,\ldots,2S\}$. Here, $\sigma$ and $2S-\sigma$ denote the numbers of
spin-1/2 singlets on the two alternating sets of bonds, and domain walls 
between different VBS domains carry effective spins.
Competition among $n$ domains with equal statistical weights produces a
permutation-symmetric infinite-randomness multicritical point
$\mathcal{P}_n$.
All $\mathcal{P}_n$ fixed points are characterized by activated energy ($\epsilon$)--length ($L$) scaling
\begin{equation}
-\ln\epsilon \sim L^{\psi_n},
\label{eq:activated}
\end{equation}
with
\begin{equation}
\psi_n = 1/n .
\label{eq:psin}
\end{equation}
This activated scaling corresponds to a divergent dynamical
exponent $z=\infty$, 
in contrast to the conventional power-law scaling $\epsilon \sim L^{-z}$ with finite $z$.
An RS$_S$ fixed point corresponds to $n=2$ with
$\psi_2=1/2$ in the Damle-Huse classification, independent of the spin
magnitude $S$. 
The spin-reduction transition between the RS$_{3/2}$ and RS$_{1/2}$ phases in
the random $S=3/2$ chain corresponds to a $\mathcal{P}_4$ fixed point, at which
the four domains $(0,3)$, $(1,2)$, $(2,1)$, and $(3,0)$ compete symmetrically,
resulting in $\psi_4=1/4$.

Earlier numerical work by Carlon \emph{et al.}~\cite{Carlon2004} employed the
density-matrix renormalization group (DMRG) to study random $S=3/2$ chains.
Their calculations did not reveal signatures of an RS phase at weak randomness
nor a $\mathcal{P}_4$ critical point. The limited system sizes, and the use of
open boundary conditions, which can introduce significant boundary effects in
the evaluation of bulk properties, may be responsible for the discrepancy
between their numerical results and the theoretical predictions.  Another
previous numerical study of random $S=3/2$ chains employed a modified SDRG
approach~\cite{Saguia2003} and also did not identify the $\mathcal{P}_4$
critical point.

In this work we revisit the \emph{spin-reduction transition} (i.e. the
$\mathcal{P}_4$ critical point) in the random spin-3/2 Heisenberg chain using
the tensor network strong-disorder renormalization group (tSDRG)
method~\cite{Goldsborough2014}.  Unlike conventional SDRG and its simple
variants, such as the approach used in Ref.~\cite{Saguia2003}, which
successively project local degrees of freedom onto the lowest-energy multiplet,
tSDRG retains multiple low-energy states at each decimation step. This extended
SDRG scheme, originally proposed in Ref.~\cite{Hikihara1999}, enables accurate
calculations of expectation values when a sufficiently large number of states
is retained during the decimation.

In our previous tSDRG study of the random $S=2$ chain with alternating bond
strength distributions~\cite{Lin2023}, we addressed the two possible phase
diagrams in the randomness ($R$)--dimerization ($D$) plane that remained
unresolved within the Damle-Huse domain-wall picture. We resolved this issue by
identifying a series of VBS phases using the disorder-averaged twist order
parameter~\cite{Nakamura2002}.  Our results support the scenario in which the transition
at $D=0$ from the gapless Haldane phase to the $\mathrm{RS}_2$ phase is
governed by a $\mathcal{P}_3$ fixed point with $\psi_3=1/3$, rather than the
alternative scenario involving a $\mathcal{P}_4$ fixed point on the undimerized
line.  This study has also revealed severe finite-size effects in disordered
higher-spin chains, demonstrating that substantially larger system sizes are
necessary to distinguish asymptotic critical scaling from crossover behavior.
The ability of tSDRG to access large system sizes therefore provides a key
advantage in this setting. These findings motivate us to apply the same method
to test the predicted $\mathcal{P}_4$ criticality in the spin-3/2 chain.

Here, we primarily focus on the undimerized case ($D=0$) and investigate the
$\mathcal{P}_4$ universality by varying the strength of bond randomness $R$.
We first derive, from the Damle-Huse RG flow equations, the theoretical
critical exponents governing the disorder-averaged bulk spin correlation
function and end-to-end correlation function at a general $\mathcal{P}_n$ fixed
point. To the best of our knowledge, theoretical predictions of these critical exponents for $n>2$ have
not been reported previously.  Our tSDRG results, including energy gap,
end-to-end correlations, and bulk correlations, overall show good agreement
with the theoretical predictions.

The paper is organized as follows.  In section~\ref{sec:RG} we review the
Damle–Huse domain-wall renormalization group and derive the critical exponents
for the disorder-averaged bulk and end-to-end correlation functions in the
$\mathcal{P}_n$ universality classes.  In section~\ref{sec:model} we define the
model and summarize the known properties of its ground-state phases. In
section~\ref{sec:method} we outline the tSDRG method. In
section~\ref{sec:results} we present our numerical results, beginning with the
energy gap and then discussing spin correlations.  We conclude in
section~\ref{sec:summary} with a summary and discussion.

\section{The Damle–Huse domain-wall renormalization group}
\label{sec:RG}

In this section, we review the Damle–Huse domain-wall renormalization group (RG) method~\cite{DamleHuse2002},
which is an extension of Fisher's SDRG scheme~\cite{Fisher1994}.

Consider the isotropic spin-$S$ Heisenberg antiferromagnetic chain with random
couplings. The Damle–Huse RG method is formulated in terms of domains  $(\sigma, 2S-\sigma)$.
A domain wall
separating domains of type-$\sigma$ and type-$\sigma'$ carries
an effective spin
\be
   \tilde{S}_{\sigma,\sigma'} = \frac{1}{2} \abs{\sigma -\sigma'}\,.  
   \label{eq:S_dw}
\ee 
The effective spin magnitude $\tilde{S}_{\sigma,\sigma'}$ is given by the
number of unpaired spin-1/2's across the wall.
Each domain is associated with a random coupling $\tilde{J}$ connecting two domain-wall spins
that bound it.

The RG can then be formulated as a sequence of domain decimations. At each
step, the domain with the largest coupling, $\tilde{J}_\text{max}=\Omega$, is
decimated.  Suppose that the decimated domain is adjacent to domains of types
$\sigma'$ on its left and $\sigma''$ on its right. Two distinct decimation
processes can occur:
\begin{itemize}
\item[(I)] Identical neighboring domains, $\sigma' = \sigma''$: 
The two domain-wall spins surrounding the decimated domain form a singlet, 
and the two neighboring domains, which have the same type, merge into a single domain with renormalized coupling
\be 
  \tilde{J}_\text{new}=\tilde{J}_\text{L}\tilde{J}_\text{R}/\Omega;
  \label{eq:Jnew}
\ee  
\item[(II)] Different neighboring domains, $\sigma' \neq \sigma''$: 
The two domain walls combine into a single new domain wall with an effective
spin $\tilde{S}_{\sigma',\sigma''}$, which now separates the adjacent domains
of types $\sigma'$ and $\sigma''$.  These two domains retain their respective
types and couplings.
\end{itemize}

\subsection{$\mathcal{P}_n$ universal properties}
\label{sec:dw_psi}
At a permutation-symmetric fixed point $\mathcal{P}_n$, all $n$ domain types
occur with equal probability $w_\sigma =1/n$, $\forall \sigma$.  For $n=2$, every decimation necessarily
involves identical neighboring domains and thus reduces to the RS phase.  For
$n>2$, both decimation branches, (I) and (II),  described above are possible.

It is more convenient to parametrize the coupling strengths on a logarithmic scale. Following Refs.~\cite{Fisher1994,DamleHuse2002,Refael2009},
we define $\Gamma=-\ln(\Omega)$ and  $\beta \equiv \ln(\Omega/\tilde{J})$, where $\Omega$ is the strength of the strongest remaining coupling.
Let $Q(\beta,\Gamma)d\beta$ denote the number of active domains  with log-coupling in the interval $[\beta, \beta+d\beta]$,
and let $\rho(\Gamma) \equiv \int_0^\beta Q d\beta$.
The probability that two neighboring domains of a decimated domain have the same type is given by $p_{\tilde{S}=0}=1/(n-1)$. 
In terms of these variables, the flow of the domain-coupling distribution can be written as
\be
   \frac{\partial Q}{\partial \Gamma} = \frac{\partial Q}{\partial \beta} - 
   2 p_{\tilde{S}=0}  \frac{Q(0,\Gamma)}{\rho(\Gamma)}  Q(\beta, \Gamma) + 
   p_{\tilde{S}=0} \frac{ Q(0,\Gamma)}{\rho(\Gamma)^2} (Q \otimes Q)(\beta)\,,
   \label{eq:flow} 
\ee
where 
\be
   (Q \otimes Q)(\beta) = \int_0^\infty \int_0^\infty d\beta_1 d\beta_2 \delta(\beta-\beta_1-\beta_2)
    Q(\beta_1,\Gamma) Q(\beta_2,\Gamma)\,.
\ee
The first term on the right-hand side describes the shift of the log-coupling $\beta$ as the RG scale $\Gamma$ increases.
The second term accounts for the removal of two neighboring domains in a same-type  event.
The last term describes the generation of a new domain 
through the merge of two same-type neighbors, with new domain coupling determined by Eq.~\eqref{eq:Jnew}.

Equation~\eqref{eq:flow} can be solved by the ansatz $Q=Q_0(\Gamma)e^{-\chi_n \beta / \Gamma}$, yielding the normalized fixed-point solution
\be
         P^*(\beta, \Gamma) = \frac{Q(\beta, \Gamma)}{\rho(\Gamma)} = \frac{\chi_n}{\Gamma} e^{-\chi_n \beta / \Gamma}\,,
         \label{eq:P_star}
\ee
with
\be
      \chi_n=\frac{1}{p_{\tilde{S}=0}}=n-1,\quad \rho(\Gamma) \sim \Gamma^{-n}\,.
       \label{eq:chi_rho}
\ee
For a finite chain of length $L$, we have $\rho(\Gamma_L)\sim 1/L$. Combined
with $\rho(\Gamma_L) \sim \Gamma_L^{-n}$, we reproduce the main result for a
permutation-symmetric multicritical point given in Eq.~\eqref{eq:psin}: $
\Gamma_L \sim L^{\psi_n}$ with $\psi_n =1/n$. 

\begin{figure}[t]
\centering
\includegraphics[width=0.6\textwidth]{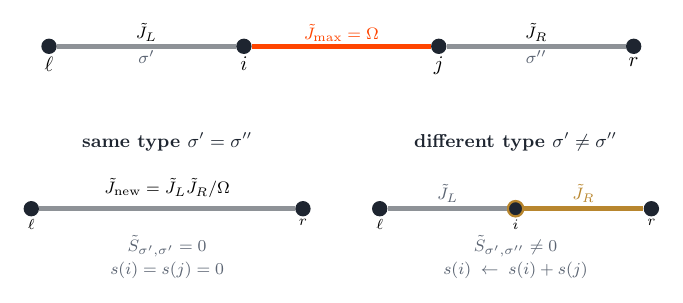}   
\caption{\label{fig:domain_rg} Two decimation processes in the Damle–Huse domain-wall RG.
In the same-type ($\sigma'=\sigma''$) process, 
the domain-wall spins at sites $i$ and $j$, which bound the domain with the strongest coupling, form a singlet, and that domain is eliminated.
In the different-type ($\sigma'\neq \sigma''$) process,
the two domain walls at $i$ and $j$ merge into a single domain wall carrying the effective spin $\tilde{S}_{\sigma',\sigma''}$
given by Eq.~\eqref{eq:S_dw}. Throughout the RG, the size $s(i)$ of spin cluster associated with each domain wall is also tracked. 
}
\end{figure}

\subsection{Universal exponents for spin correlations at $\mathcal{P}_n$}
\label{sec:dw_eta}

To provide theoretical results for spin correlations, below we explicitly
derive the critical exponents associated with the disorder-averaged spin correlation functions.

At the  $\mathcal{P}_n$ critical point, the disorder-averaged spin-spin correlation
$\overline{C}(i,j)= \overline{\e{\mathbf{S}_i\cdot \mathbf{S}_j}}$ (overline denotes disorder average and $\mathbf{S}_i$ represents the
spin-$S$ operator at site $i$) decays algebraically,
\begin{align}
\text{bulk:} \qquad & \overline{C}_b(r) \sim r^{-\eta}, \qquad r=\abs{i-j}, \\
\text{end-to-end:} \qquad & \overline{C}_1(L) \equiv \overline{C}(1,L) \sim L^{-\eta_1},
\end{align}
defining two universal critical exponents, $\eta$ and $\eta_1$, for bulk and end-to-end correlations, respectively.
Here and throughout the paper, site indices follow the one-based convention, starting from $1$.

Within the SDRG picture, the leading contribution to the disorder-averaged
spin-spin correlation is determined by strongly correlated singlet pairs~\cite{Fisher1994}.
Thus, the correlation between two marked sites is proportional to 
the probability that the two sites are decimated into the same singlet,
\be
        C(i,j) \sim P_\text{pair}(i,j)\equiv \text{Pr}[i,j\, \text{ decimated into the same singlet}]
\ee
At the $\mathcal{P}_2$ fixed point ($n=2$), the exponents $\eta=2$ and $\eta_1=1$ are known from Fisher's 
SDRG analysis~\cite{Fisher1994,FisherYoung1998}. 
Here we extend the derivation of $\eta$ and $\eta_1$ to the general $\mathcal{P}_n$ fixed points
within the domain-wall framework.

To characterize the spatial extent of the effective spin clusters, we assign to
each active domain wall $i$ an integer cluster size $s(i)\geq 1$, with $s(i)=1$
for every wall initially. In the decimation channel with identical neighboring
domains,  $\sigma'=\sigma''$, the two domain walls, $i$ and $j$, bounding
the domain with the strongest coupling are removed and replaced by a renormalized long-range
coupling. In the decimation channel with different neighboring domains,
$\sigma'\neq\sigma''$, one domain wall, say $i$, survives and absorbs the
cluster associated with the other wall, $j$, according to
\be
   s(i) \;\leftarrow\; s(i) + s(j).
   \label{eq:cluster_merge}
\ee
Let $\Pi(s, \Gamma)$ denote the normalized probability that an active domain wall at RG energy scale $\Gamma$ 
has a cluster of size $s$. Thus, $\rho(\Gamma)\Pi(s,\Gamma)$ gives the density of active clusters of size $s$ at scale $\Gamma$. 
The flow equation for the cluster-size distribution is
\be
     \frac{\partial}{\partial \Gamma} \left[\rho(\Gamma) \Pi(s, \Gamma) \right]
     = -2Q(0, \Gamma)  \Pi(s, \Gamma) + p_{\tilde{S} \neq 0} Q(0, \Gamma) \sum_{s_1, s_2} \delta_{s_1+s_2, s} 
        \Pi(s_1) \Pi(s_2)\,,  
     \label{eq:cluster_flow}
\ee 
where $p_{\tilde{S}\neq 0}=1-p_{\tilde{S}=0}=(n-2)/(n-1)$.
The first term on the right-hand side of Eq.~\eqref{eq:cluster_flow} accounts for the decimation of two active
clusters of size $s$, giving a factor of two because two domain walls are
involved in each decimation event. The second term describes the formation of a
new cluster of size $s_1+s_2$ through the cluster-merging process in Eq.~\eqref{eq:cluster_merge}.
By changing the RG variable from $\Gamma$ to $\rho(\Gamma)$ and using $d\rho/d\Gamma = -n\rho/\Gamma$, we obtain
\be
  \rho \frac{\partial \Pi(s, \rho)}{\partial \rho} 
    = \left( 1-\frac{2}{n} \right)
    \left[ \Pi(s, \rho) - \sum_{s_1 =1}^{s-1} \Pi(s_1, \rho) \Pi(s-s_1, \rho)\right]\,.
  \label{eq:flow_cluster} 
\ee
Interestingly, this equation is algebraically identical to the cluster-size flow equation for the SU($n$) chain derived in Ref.~\cite{Hoyos2004}.
Following Ref.~\cite{Hoyos2004}, we solve Eq.~\eqref{eq:flow_cluster} using the ansatz
$\Pi(s,\rho) = \rho^\gamma \exp(-s \rho^\gamma)$ with $\gamma\equiv 1-2/n$,
obtaining the mean cluster size $\overline{s}_\rho = \int s \Pi(s,\rho) ds = \rho^{-\gamma}$.
The fraction of the original sites that remain within some active clusters is therefore
$\rho \overline{s}_\rho=\rho^{1-\gamma}$.
For two bulk sites $i$ and $j$ to be decimated into the same singlet, both
sites must remain within active clusters up to the RG scale at which they are
paired. To leading order, the corresponding pairing probability is therefore
$P_\text{pair}(i,j) \sim [\rho(\Gamma)^{1-\gamma}]^2 \sim \Gamma^{-4}$.
Using the activated energy-length relation $\Gamma\sim r^{\psi_n}$ at the $\mathcal{P}_n$ fixed point, we obtain
\be
      \overline{C}_b(r) \sim \Gamma^{-4} \sim r^{-4\psi_n}\,,
\ee 
and hence the $n$-dependent exponent
\be 
    \eta(n)=4\psi_n=\frac{4}{n}\,,
    \label{eq:Cb_eta}  
\ee
for the disorder-averaged bulk spin correlation function.

Next, we derive the critical exponent $\eta_1$ for disorder-averaged
end-to-end correlations in a chain with open boundary conditions. Let $\ell_1$
denote the original leftmost site, and let $f(\Gamma)$ be the probability that
$\ell_1$ remains active at the RG energy scale $\Gamma$. 
When the domain containing $\ell_1$ is selected for decimation, corresponding to the strongest
remaining coupling $\beta=0$, the site $\ell_1$ is decimated with probability
$p_{\tilde{S}=0}$ and survives with probability $1-p_{\tilde{S}=0}$. 
Here we assume that the end domain has the same decimation rate as any interior domain. 
Thus, we obtain
\be
              \frac{df(\Gamma)}{d\Gamma} = -p_{\tilde{S}=0} P^*(0,\Gamma) f(\Gamma) = - \frac{f(\Gamma)}{\Gamma}\,,
\ee
where we have used $P^*(0,\Gamma)=\chi_n/\Gamma$ (Eq.~\eqref{eq:P_star}) and $\chi_n p_{\tilde{S}=0}=1$ (Eq.~\eqref{eq:chi_rho}).
Solving this equation gives $f(\Gamma) \sim 1/\Gamma$. The probability $f(\Gamma)$ is also for
the original rightmost site $r_{L}$ remaining active at the RG scale $\Gamma$.

The disorder-averaged end-to-end correlation is governed by the probability
that the two end spins are ultimately paired into a singlet. The corresponding
pairing probability is
\be
                \overline{C}_1(L) \sim [f(\Gamma)]^{2} \sim \Gamma^{-2} \sim L^{-2 \psi_n}\,.
\ee
which yields the critical exponent 
\be 
     \eta_1(n) = 2\psi_n = \frac{2}{n}\,.
     \label{eq:C1_eta}
\ee

\subsection{Numerical implementation of the domain-wall RG}
\label{sec:mc_rg}

\begin{figure}[t]
\centering
\includegraphics[width=0.9\textwidth]{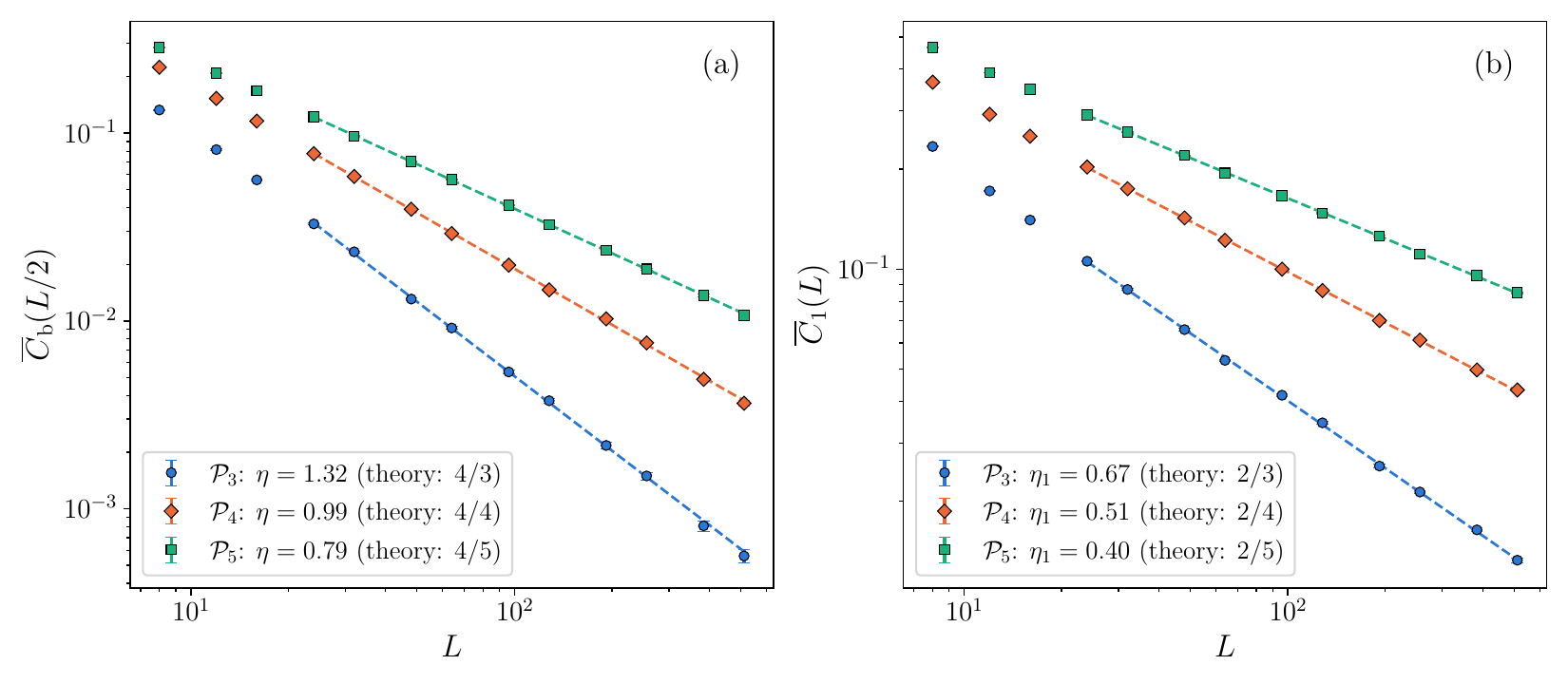} 
\caption{\label{fig:nrg_C} Disorder-averaged correlations at the $\mathcal{P}_n$ fixed point, with $n=3, 4$ and $5$, 
obtained by numerical domain-wall RG. (a) Disorder-averaged bulk correlations for spins separated by the largest distance $L/2$ in
chains with periodic boundary conditions; (b) disorder-averaged end-to-end correlations in open chains.
For each parameter $n$ and system size $L$, 300,000 samples (disorder realizations) were used to obtain the averaged correlations.}
\end{figure}

In the previous subsection, we derived the critical exponents $\eta$ and
$\eta_1$ at the $\mathcal{P}_n$ fixed point by solving the cluster-size flow
equations. The Damle--Huse RG scheme can also be implemented numerically. The
procedure is as follows. We first generate an initial sequence of domains by
assigning each of the $n$ domain types an equal probability $1/n$, subject to
the constraint that no two adjacent domains have the same type. This
construction corresponds to the $\mathcal{P}_n$
fixed-point ensemble. For chains with open boundary conditions, the
probability of singlet formation for a boundary domain, which has only one
neighboring domain, is assumed to be $p_{\tilde{S}=0}=1/(n-1)$, i.e., the same as that for an
interior domain. The initial distribution of domain couplings does not affect
the asymptotic fixed-point behavior; accordingly, we choose a uniform
distribution over $(0,1]$ in our simulations.

Given $M$ independent and identically distributed realizations, 
the disorder-averaged correlation between the two marked sites is estimated as
\be
         \overline{C}(i,j) = \frac{1}{M} \sum_{m=1}^M \mathds{1}_{\tilde{S}=0}(i,j)\,,
         \label{eq:mc_rg}
\ee  
where $\mathds{1}_{\tilde{S}=0}(i,j)=1$ if $i$ and $j$ are eventually paired into a same singlet
and $\mathds{1}_{\tilde{S}=0}(i,j)=0$ otherwise.
Importantly, if site $i$ is first incorporated into a cluster and site $j$ is 
independently merged into another cluster, a subsequent singlet decimation between these two clusters also contributes $1$
to the estimator $\mathds{1}_{\tilde{S}=0}(i,j)$.

Figure~\ref{fig:nrg_C} presents the numerical RG results for the
disorder-averaged bulk and end-to-end correlation functions for $n=3,4$, and
$5$. For the bulk correlations $\overline{C}_b$, we consider the correlations
between sites separated by the largest distance $\abs{i-j}=L/2$, in periodic chains of
various system sizes up to $L=512$.  For the end-to-end correlations
$\overline{C}_1$, we consider the two boundary sites, $i=1$ and $j=L$, in
open chains.  The numerical results for both $\overline{C}_b$ and
$\overline{C}_1$ reproduce the theoretical results derived above.  

We note that previous numerical studies of spin-1 and spin-2 chains using DMRG and tSDRG~\cite{Lajko2005,Tsai2020,Lin2023} also reported
estimated end-to-end correlation exponents of $\eta_1\approx 0.67-0.7$ for $\mathcal{P}_3$, in reasonable agreement with the theoretical
prediction $\eta_1=2/3$, providing independent support for this result.


\section{The spin-$3/2$ chain}
\label{sec:model}

We consider an even-length spin-3/2 antiferromagnetic Heisenberg chain with
nearest-neighbor interactions,
\begin{equation}
H = \sum_i J_i \, \mathbf{S}_i \cdot \mathbf{S}_{i+1},
\label{eq:H}
\end{equation}
where $\mathbf{S}_i$ is the spin-3/2 operator at site $i$. For a chain of $L$
sites, the sum runs from $i=1$ to $L-1$ for open boundary conditions (OBC),
and from $i=1$ to $L$ for periodic boundary conditions (PBC), with
$\mathbf{S}_{L+1}=\mathbf{S}_1$. We use OBC to study end-to-end correlations
and PBC to study bulk correlations and energy gaps. 

\begin{figure}[t]
\centering
\includegraphics[width=0.4\textwidth]{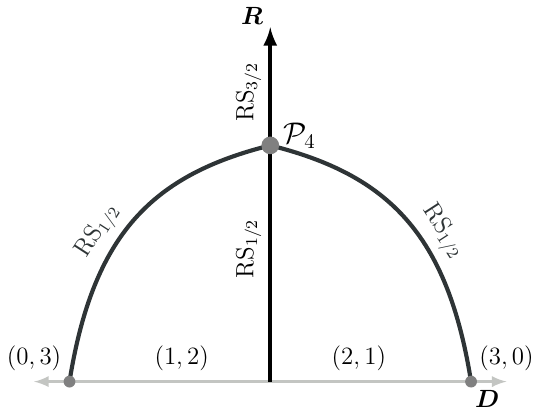} 
\caption{\label{fig:phase} Schematic phase diagrams for the spin-3/2 antiferromagnetic chains in the $R$-$D$ plane.}
\end{figure}

We first summarize the established ground-state properties of the random
spin-3/2 chain.  By introducing bond randomness (parameterized by $R$) and
dimerization (parameterized by $D$) into the couplings, Damle and
Huse~\cite{DamleHuse2002} proposed a possible $R$-$D$ phase diagram for the
spin-3/2 chain, as schematically illustrated in Fig.~\ref{fig:phase}. A
representative coupling distribution is $J_i=J_0[1+(-1)^i D]\exp(R \xi_i)$,
where the $\xi_i$ random numbers with zero mean, unit variance, and finite
moments of all orders.  Nonzero dimerization breaks the statistical equivalence
of even and odd bonds, thereby selecting among the VBS phases $(0,3)$, $(1,2)$,
$(2,1)$, and $(3,0)$.

For $R>0$, all phase boundaries between these VBS phases in the $R$-$D$ phase
diagram belong to the permutation-symmetric $\mathcal{P}_n$ universality
classes~\cite{DamleHuse2002}, governed by infinite-randomness fixed points.
These fixed points are characterized by the activated scaling exponent
$\psi_n=1/n$, the correlation-length critical exponent $\nu_n=2 \psi_n^{-1}/
(\sqrt{4n+1}-1)$~\cite{DamleHuse2002}, and the critical exponents governing
bulk and end-to-end correlations, given in Eqs.~\eqref{eq:Cb_eta} and \eqref{eq:C1_eta}, respectively.

At $R=0$ and $D=0$, the uniform spin-3/2 chain is gapless, with low-energy
physics described by the level-1 SU(2) Wess--Zumino--Witten
model~\cite{AffleckHaldane1987}.  Along $D=0$, the spin-reduction scenario
predicts two disorder-controlled regimes separated by a critical randomness
strength $R_c$~\cite{Refael2002,DamleHuse2002}. For $0<R<R_c$, the low-energy
degrees of freedom are effective spin-1/2 moments associated with domain walls
between the $(1,2)$ and $(2,1)$ VBS domains. These moments form the
$\mathrm{RS}_{1/2}$ phase. For $R>R_c$, the full spin-3/2 moments form random
singlets, giving rise to the $\mathrm{RS}_{3/2}$ phase. Both regimes belong to
$\mathcal{P}_2$ universality class and exhibit activated energy--length scaling
with $\psi=1/2$, despite their different effective spin content.

At $R=R_c$ and $D=0$, all four domains $(0,3)$, $(1,2)$, $(2,1)$, and
$(3,0)$ occur with equal statistical weights at the renormalization-group
fixed point. This defines the multicritical point in the $\mathcal{P}_4$ universality class, characterized by
\begin{equation}
\psi_4 = \frac{1}{4}, \qquad
\eta = 1, \qquad
\eta_1 = \frac{1}{2}\,.
\end{equation}
The permutation symmetry at $R_c$ is emergent,
as it is not a symmetry of the bare Hamiltonian.

In our numerical study, we focus on the critical line at $D=0$.
We consider a power-law distribution of the random couplings, 
\be
        P(J) = 
\begin{cases}
R^{-1} J^{-1+1/R}, & 0 < J \leqslant 1, \\
0, & \text{otherwise},
\end{cases}
\label{eq:PJ}
\ee
with $R>0$.  We set the upper cutoff of the coupling to unity.  The variable
$-\ln J_i$ follows an exponential distribution with mean and standard deviation
$R$; hence $R$ measures the width of the logarithmic coupling distribution.
Increasing $R$ broadens the distribution, while the clean limit $R\to 0$ gives
$J_i=1,\,\forall i$. This parametrization has been widely used in numerical
studies of disordered spin chains~\cite{Lajko2005,Tsai2020,Lin2023,Carlon2004}.
Our goal is to numerically locate $R_c$ in the spin-3/2 chain and characterize
the critical scaling at this multicritical point.

\section{The numerical method}
\label{sec:method}


To numerically test the predictions of the Damle--Huse domain-wall RG, we
perform first-principles calculations of the spin-$3/2$ chain using tSDRG, a
tree-tensor-network generalization of SDRG that retains multiple low-energy
states at each renormalization step, rather than only the local ground
state~\cite{Hikihara1999,Goldsborough2014}.  The method has previously been
applied to random quantum Ising chains~\cite{Lin2017} and random spin-1 and
spin-2 Heisenberg chains~\cite{Tsai2020,Lin2023}. 
Here, we summarize the main steps below.

Starting from a matrix product operator (MPO) representation of the
Hamiltonian, we iteratively merge adjacent blocks, each initially comprising two sites,
according to their local energy spectra.  At each step, we select the pair of
blocks with the largest energy gap separating the states to be retained from
those to be discarded, rather than simply choosing the pair connected by the
largest bare coupling~\cite{Lin2023}. The selected pair is replaced by a
renormalized block containing at most $\chi$ low-energy states. The number of
retained states is adjusted to preserve complete SU(2) multiplets.  The
corresponding retained eigenvectors define an isometric tensor that projects
the MPO and the relevant operators onto the reduced Hilbert space.  The
Hamiltonians of the neighboring blocks are then updated, and the procedure is
repeated until only two renormalized blocks remain.

Diagonalizing the final, or top, Hamiltonian yields the approximate ground
state, represented by an inhomogeneous binary tree tensor network. Spin
correlations are evaluated by contracting the corresponding operators with
this network; the isometric structure makes these contractions efficient.
For each disorder realization, we estimate the energy gap from the difference
between the lowest excited-state and ground-state energies of the final
Hamiltonian. Further algorithmic details, including the MPO construction
and the isometric tensors, are given in~\cite{Goldsborough2014,Lin2023}.

We use a maximum retained dimension $\chi=40$. 
System sizes up to  $L=1280$ are considered.
At each parameter point, observables are averaged over approximately
$10^4$--$3\times 10^4$ disorder realizations. Access to these large systems
is important for resolving the finite-size crossover effects discussed in
the introduction. The results nevertheless remain subject to finite-size
and finite-$\chi$ effects, which must be distinguished from asymptotic
critical behavior.

\section{Numerical results}
\label{sec:results}

Our key finding is the reentrant behavior of the RS phase in the undimerized
random spin-$3/2$ chain as the bond-randomness strength $R$ increases, together
with the existence of a $\mathcal{P}_4$ multicritical point at $R_c\approx
0.9$. In the following, we present numerical results supporting these findings.

\subsection{Energy gap}
\label{sec:gap}

\begin{figure}[t]
\centering
\includegraphics[width=0.9\textwidth]{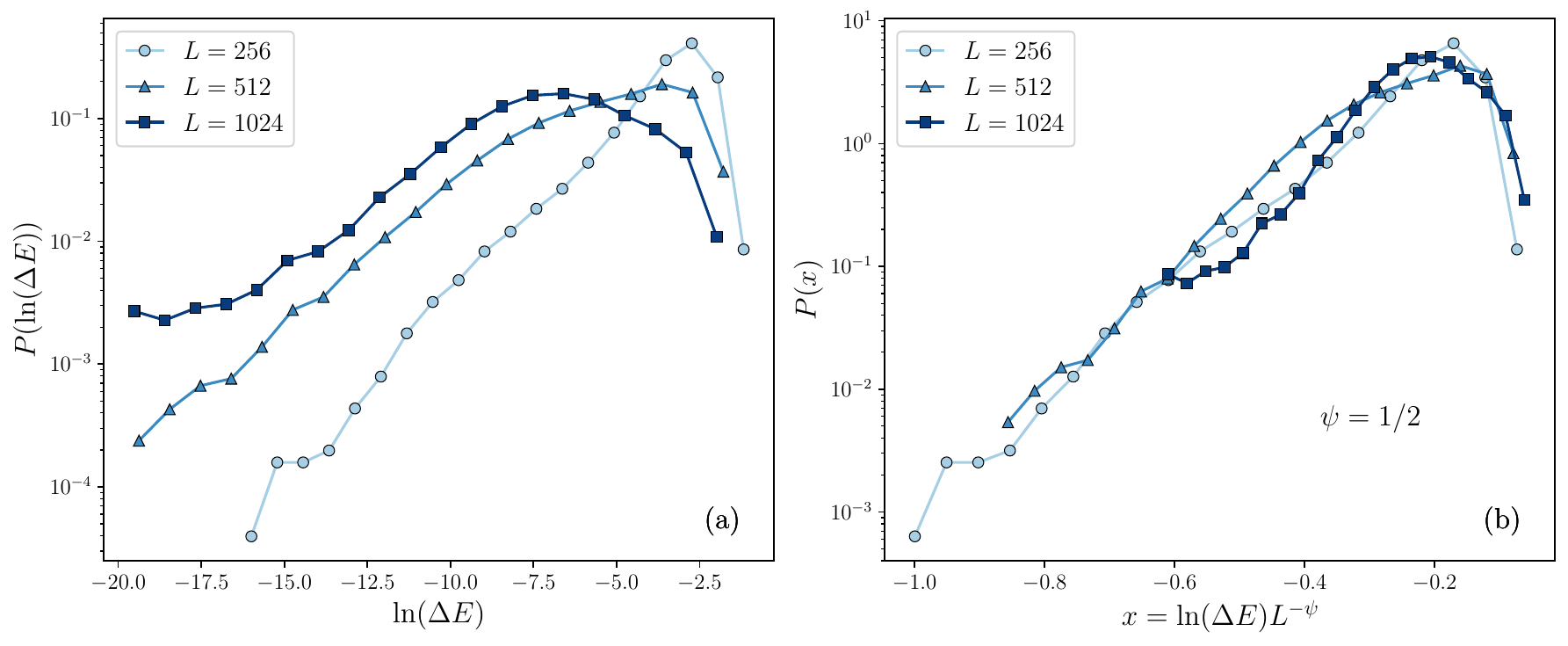} 
\caption{\label{fig:gap_R010} Distributions of energy gaps at $R=0.1$. (a) Unscaled log-histograms for system sizes $L=256, 512$, and $1024$; 
(b) A scaling plot of the gap-histograms, assuming $-\ln(\Delta E) \sim L^{\psi}$ with $\psi=1/2$.}
\end{figure}

For each disorder realization of a periodic chain, we estimate the energy gap
$\Delta E$ from the difference between the lowest excited-state and
ground-state energies of the effective Hamiltonian obtained at the final RG
step.  At an infinite-randomness fixed point, the energy gaps obey activated
energy--length scaling  [Eq.~\eqref{eq:activated}] and exhibit an extremely
broad distribution that becomes increasingly broad with system size, even on a
logarithmic scale.  When expressed in terms of the rescaled logarithmic
variable $-\ln(\Delta E)/L^{\psi_n}$, the gap distributions for different
system sizes $L$ are expected to collapse onto a size-independent limiting
function. In principle, this scaling collapse provides a means of verifying
activated scaling and extracting the associated exponent $\psi_n$.

However, unlike conventional SDRG, tSDRG cannot be formulated directly in terms
of the logarithmic energy scale. At an infinite-randomness fixed point, the
exponentially small energy gaps can also cause the eigensolver to fail to
converge accurately, leading to noisy gap estimates. Similar numerical
difficulties have been reported in previous
studies~\cite{Lin2017,Igloi2012,Zhao2016,Roosz2017}.  Nevertheless, the log-gap
histogram for weak randomness at $R=0.1$, shown in Fig.~\ref{fig:gap_R010},
exhibits a clear signature of activated scaling and agrees well with the
expected exponent $\psi=1/2$, supporting the existence of an RS phase at weak
randomness.

\begin{figure}[t]
\centering
\includegraphics[width=0.5\textwidth]{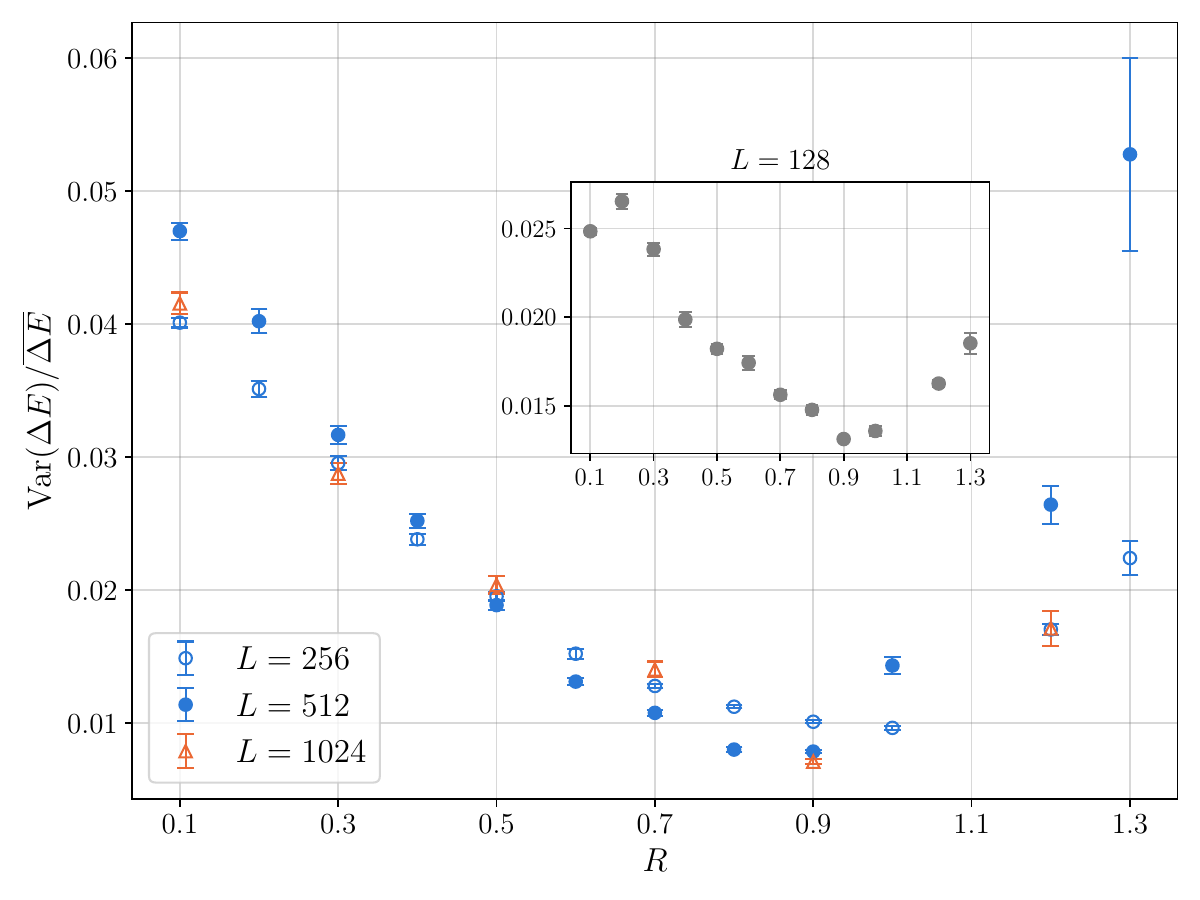} 
\caption{\label{fig:B} The ratio ($B$) of the variance of the energy gap to the mean energy gap, plotted as a function of $R$
for system sizes $L=128 (\text{inset}), 256, 512$, and $1024$. The minimum around $R=0.9$ indicates a transition point.}
\end{figure}

Rather than extracting the exponent $\psi_n$ at various values of $R$ through
data collapse, we develop a complementary procedure to detect the anticipated
reentrant transition to the RS phase as the bond-randomness strength $R$
increases and to locate the critical value $R_c$ associated with the
$\mathcal{P}_4$ fixed point. We define the ratio
\be
        B(R) \equiv \frac{\text{Var}[\Delta E]}{\overline{\Delta E}}\,.
        \label{eq:B}
\ee
Using the fixed-point solution in Eq.~\eqref{eq:P_star} and identifying the logarithmic gap 
$\beta = -\ln \Delta E$ at the final RG scale $\Gamma_f(R,L)$ for a finite chain of length $L$,
we obtain, to the leading order,
\be
     \overline{\Delta E} = \frac{\lambda}{1+\lambda},\quad \text{Var}(\Delta E) = \frac{\lambda}{(2+\lambda)(1+\lambda)^2}\,,
     \label{eq:mean_var}
\ee
where $\lambda\equiv\chi_n/\Gamma_f$. Thus,
\be
       B(R) = \frac{1}{(1+\lambda)(2+\lambda)}\,.
\ee
At a $\mathcal{P}_n$ fixed point, $\lambda$ is expected to scale as $\lambda
\sim \chi_n L^{-\psi_n}$.  Using the known fixed-point values $(\chi_2=1,
\psi_2 = 1/2)$ for the RS phase and $(\chi_4=3, \psi_4 = 1/4)$ for the
$\mathcal{P}_4$ fixed point, we expect $B_{\mathcal{P}_4}<B_{\mathrm{RS}}$.

It is worth noting that the variance $\text{Var}(\Delta E)$ of a function of
$R$ does not exhibit a clear minimum in our numerical results. We explain this
behavior as follows.  When expressed in terms of $\lambda$
[Eq.~\eqref{eq:mean_var}], the variance $\text{Var}(\Delta E)$ exhibits a
maximum at $\lambda^*=(\sqrt{5}-1)/2$. At the same time,
$\lambda(R)=\chi/\Gamma_f$ exhibits a bump around $R_c$, reflecting the change
in the scaling exponent from $\psi_2$ to $\psi_4$.  The nonmonotonic dependence
of the variance on $\lambda$, combined with the bump in $\lambda(R)$, makes
$\text{Var}(\Delta E)$ alone an unreliable indicator for locating $R_c$.  By
normalizing the variance by the mean energy gap $\overline{\Delta E}$, we
obtain the ratio $B(R)$, which varies monotonically with $\lambda$.  This
normalization transforms the bump in $\lambda(R)$ near $R_c$ into a clear
minimum in $B(R)$, providing a more robust indicator of the $\mathcal{P}_4$
critical point.

\begin{figure}[t]
\centering
\includegraphics[width=0.95\textwidth]{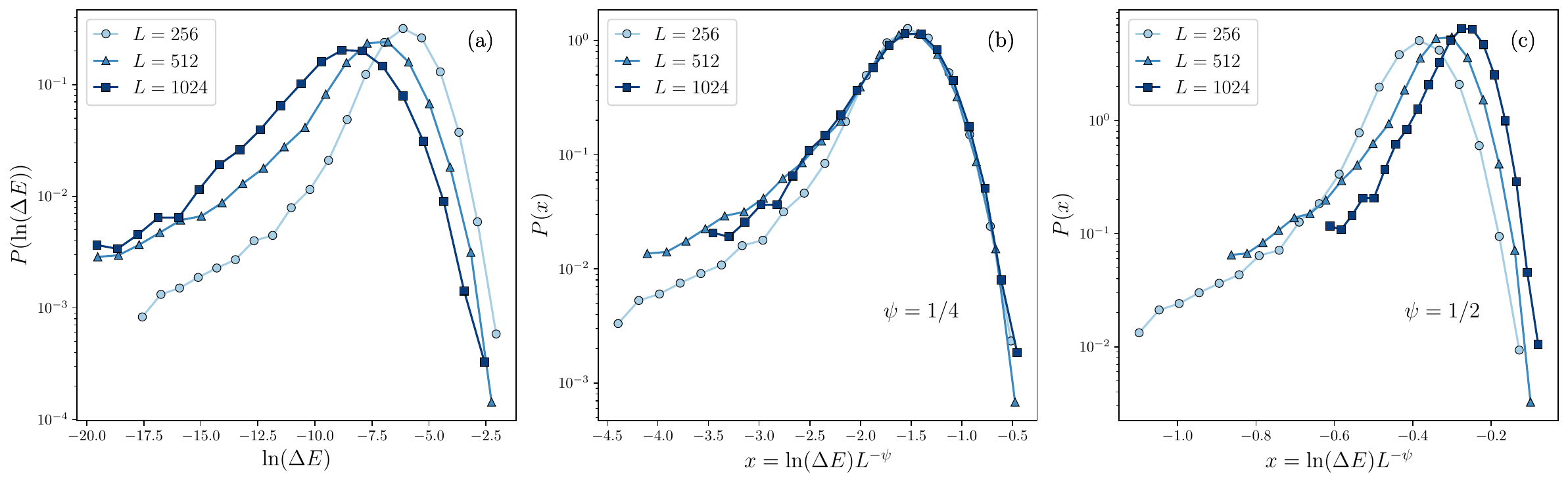} 
\caption{\label{fig:gap_R090} Distributions of energy gaps at $R_c=0.9$. (a) Unscaled log-histograms for $L=256, 512$, and $1024$; (b) 
A scaling plot of the gap-histograms with $\psi_4=1/4$, corresponding to the $\mathcal{P}_4$ universality class; 
(c) Rescaled histograms with $\psi_2=1/2$ exhibit poor data collapse.}
\end{figure}

Figure~\ref{fig:B} shows the ratio $B(R)$ as a function of $R$ for periodic chains of lengths $L=128, 256, 512$, and $1024$.
The finite-size ratios vary smoothly with $R$ and exhibit a minimum around $R\approx0.9$, 
suggesting a reentrant transition to the RS phase and locating the $\mathcal{P}_4$ critical point at approximately $R_c\approx0.9$.
The gap distribution at $R_c=0.9$ and the corresponding scaling plots in Fig.~\ref{fig:gap_R090} 
provide further support for the $\mathcal{P}_4$ scaling.

\subsection{Correlations}
\label{sec:corr}

We now turn to the tSDRG results for the spin correlation functions.
For chains of even length $L$ with OBC, we measure the end-to-end correlation,
\be
         C_1(L) = \e{\mathbf{S}_1 \cdot \mathbf{S}_{L}}\,. 
\ee
For bulk correlations, we consider periodic chains and examine the longest-distance correlations at  $r=L/2$ as a function of the system size $L$.
For each disorder realization at fixed system size $L$, we average the bulk correlations over all pairs of spins $(i,i+L/2)$ in the chain:
\be
            C_b(L/2) = \frac{2}{L}\sum_{i=1}^{L/2} \e{\mathbf{S}_i \cdot \mathbf{S}_{i+L/2}}\,.
\ee

\begin{figure}[t]
\centering
\includegraphics[width=0.5\textwidth]{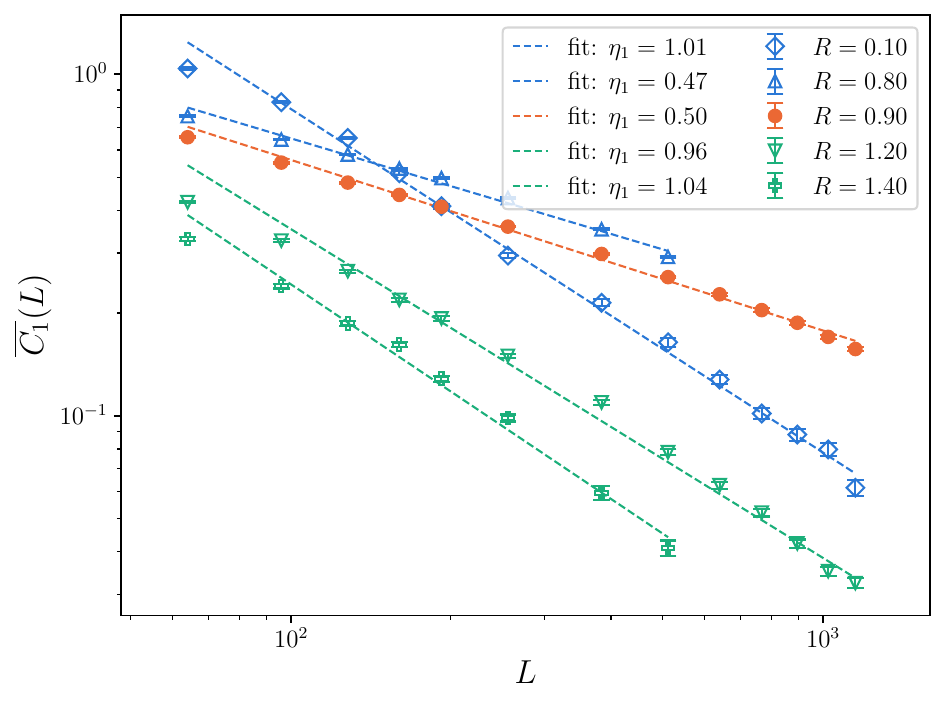} 
\caption{\label{fig:C1_all} Disorder-averaged end-to-end correlation functions for various  bond-randomness strengths $R$,
ranging from weak randomness $R=0.1$ to strong randomness $R=1.4$.
For both weak randomness $R=0.1$ and strong randomness $R=1.2$ and $R=1.4$,
the fitted critical exponents approach the theoretical value $\eta_1 =1$ for the RS phase.
At the $\mathcal{P}_4$ critical point, estimated to occur at $R_c=0.9$,
the fitted exponent shows excellent agreement with the theoretical prediction $\eta_1=1/2$, derived in Sec.~\ref{sec:dw_eta}.}
\end{figure}

\begin{figure}[t]
\centering
\includegraphics[width=0.5\textwidth]{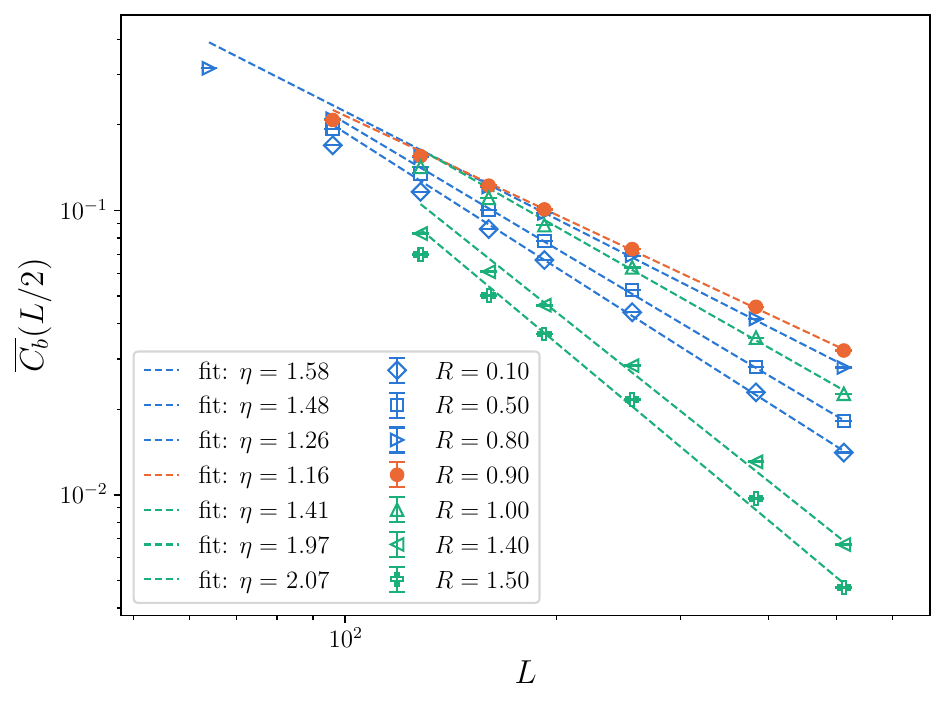} 
\caption{\label{fig:Cb_all} Disorder-averaged bulk correlation functions at the largest distance $r=L/2$ for various $R$,
ranging from weak bare randomness $R=0.1$ to strong randomness $R=1.5$.
At the estimated $\mathcal{P}_4$ critical point $R_c=0.9$, the fitted exponent $\eta\approx 1.16$
deviates slightly from the theoretical prediction  $\eta=1$. For $R>R_c$, the fitted exponents at  $R=1.4$ and $R=1.5$
agree well with the $\mathcal{P}_2$ universality prediction $\eta\approx 2$.
At weak randomness, the approach to $\mathcal{P}_2$ universality appears to be considerably slower.
Nevertheless, the nonmonotonic dependence of the fitted exponent on $R$ indicates reentrant behavior as the randomness increases.}
\end{figure}

Figures~\ref{fig:C1_all} and \ref{fig:Cb_all} show the disorder-averaged end-to-end
and bulk correlation functions, respectively, for various bond-randomness
strengths $R$.  In both cases, the estimated critical exponents, extracted from
the power-law decay of the correlations in the log-log plots, exhibit a
nonmonotonic dependence on $R$, indicating reentrant behavior as the randomness
increases. 
At the $\mathcal{P}_4$ critical point at $R_c=0.9$, we estimate
$\eta_1\approx 0.5$ for the disorder-averaged end-to-end correlation and
$\eta\approx 1.16$ for the bulk correlation.  Both estimates are in good
agreement with the theoretical predictions $\eta_1=1/2$ and $\eta=1$,
respectively.
It is worth noting that the estimated bulk correlation exponent $\eta$
reaches a minimum at $R_c \approx 0.9$, providing an additional means of
identifying the $\mathcal{P}_4$ multicritical point. In contrast, the estimated end-to-end
correlation exponent $\eta_1$ changes to $\eta_1< 0.5$ in the intermediate-disorder regime, which
may reflect quasi-surface order, as observed in
Refs.~\cite{Carlon2004,Fath2006}.

\begin{figure}[t]
\centering
\includegraphics[width=0.95\textwidth]{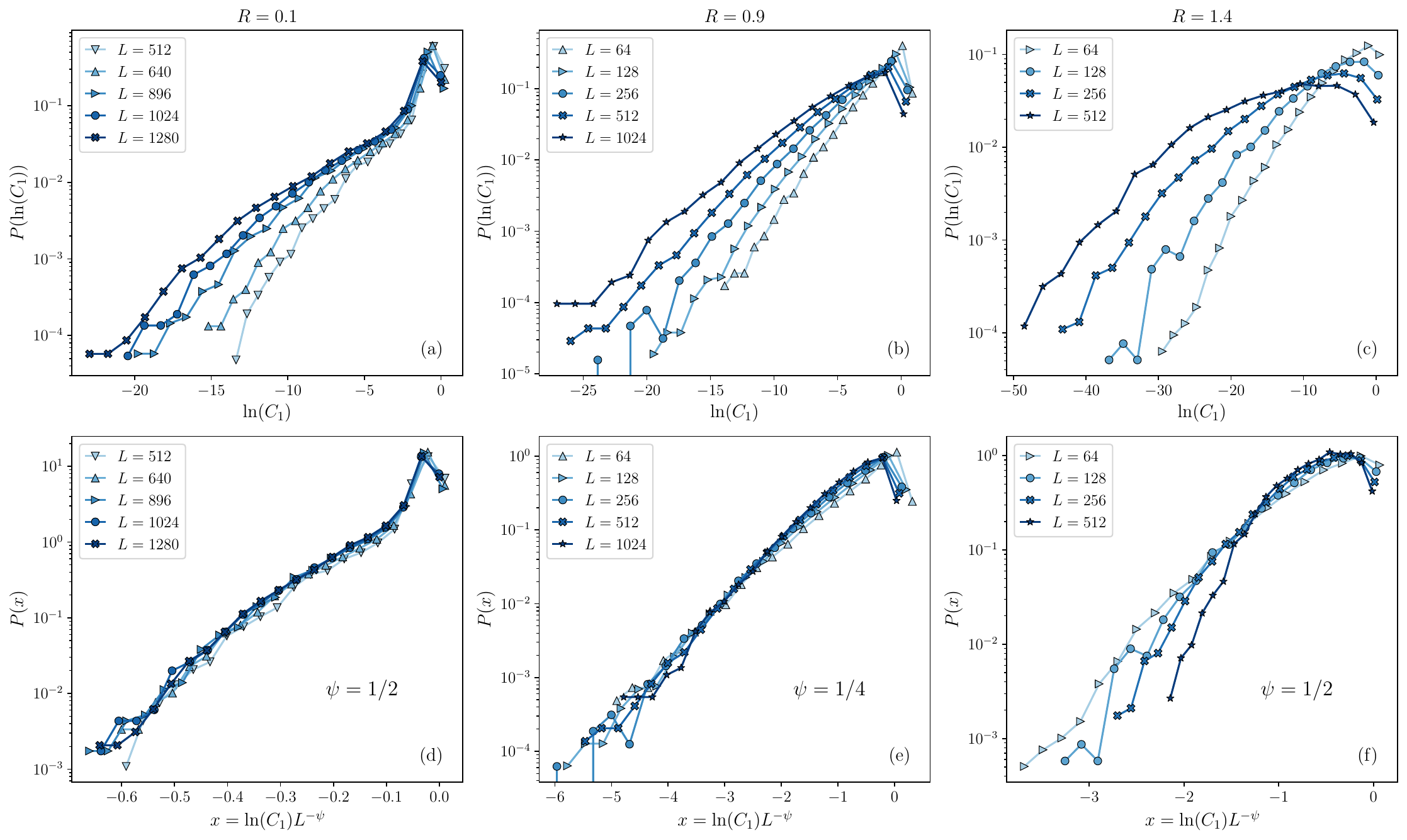}
\caption{\label{fig:C1_histo} Distributions of end-to-end correlations at $R=0.1, 0.9$ and $1.4$. The upper panels show the unscaled distributions, while
the lower panels show the corresponding scaling plots in terms of the scaled variable $x=\ln(C_1)L^{-\psi}$, using 
$\psi=1/2$ for $R=0.1$ and $R=1.4$, $\psi=1/4$ at $R_c=0.9$.}
\end{figure}

We also examine the full distribution of $C_1$. End-to-end correlations probe
the broad low-energy fluctuations characteristic of infinite-randomness physics~\cite{FisherYoung1998,Igloi1998}. 
Their characteristic logarithmic scale grows as
\begin{equation}
-\ln C_1(L) \sim L^{\psi}\,,
\label{eq:lnC1}
\end{equation}
and thus the distribution of $\ln C_1$ broadens with increasing system size $L$.
Using the rescaled variable $x=\ln(C_1)/L^{\psi}$, where $\psi$ is the corresponding scaling exponent, 
we can test whether the finite-size distributions collapse onto a size-independent limiting
distribution. Figure~\ref{fig:C1_histo} show the
unscaled distributions of  $\ln C_1$ and the corresponding scaling plots for $R=0.1, 0.9$, and $1.4$.
In all three cases, the unscaled distributions broaden with increasing  $L$, although
larger system sizes are required to observe this trend clearly in the weak-randomness case ($R=0.1$).
The scaling plots exhibit good data collapse for $R=0.1$ and at the $\mathcal{P}_4$ multicritical point ($R_c=0.9$),
using $\psi=1/2$ and $\psi=1/4$, respectively. For $R=1.4$, the scaling
with $\psi=1/2$ collapses the distributions near their peaks, but a
substantial system-size dependence persists in the negative tail.
This deviation at strong randomness may arise from numerical inaccuracies in
evaluating exponentially small correlation values.

\section{Summary and discussion}
\label{sec:summary}

We have revisited the spin-reduction transition in the random spin-3/2
antiferromagnetic Heisenberg chain using tSDRG, focusing on the undimerized
line. A central challenge is that the $\mathrm{RS}_{1/2}$ and
$\mathrm{RS}_{3/2}$ phases share the same RS scaling
exponents despite their different effective spin content. Thus, these
exponents alone cannot distinguish the two phases. Instead, we identify the
critical point separating the two regimes through its distinct critical
scaling behavior. Our results support the spin-reduction scenario proposed by
Refael, Kehrein, and Fisher~\cite{Refael2002} and its interpretation in terms of the
permutation-symmetric $\mathcal{P}_4$ fixed point of Damle and Huse~\cite{DamleHuse2002}.

A key piece of evidence comes from the energy-gap distribution. We construct a
diagnostic based on the ratio of the gap variance to the mean gap, which
identifies the critical randomness $R_c \approx 0.9$ associated with the
$\mathcal{P}_4$ fixed point. 
At $R_c$, the distributions of the logarithmic energy gap, $\ln \Delta E$, for chains of up to $L=1024$ 
collapse over a broad range when rescaled by $L^{1/4}$, consistent with the predicted activated scaling exponent
$\psi_4=1/4$. At weak randomness $R=0.1$, the finite-size scaling of the energy-gap distributions
likewise supports an RS phase with $\psi=1/2$.

Spin correlations provide further evidence for the $\mathcal{P}_4$ fixed
point and the reentrant behavior of the RS phase as the bond-randomness
strength $R$ varies. 
At $R_c$, the estimated critical exponents associated with
the disorder-averaged end-to-end and bulk correlation functions are
$\eta_1\approx 0.5$ and $\eta\approx 1.16$, respectively. These estimates
are consistent with the theoretical predictions $\eta_1=1/2$ and $\eta=1$
for the $\mathcal{P}_4$ universality class, which we derive in this work.
Furthermore, the $R$-dependence of the estimated exponents $\eta_1$ and
$\eta$, together with the finite-size scaling of the $\ln C_1$
distributions, provides additional evidence for the reentrant RS transition.

The ability of tSDRG to access large system sizes is a significant advantage in
the study of disordered systems. Nevertheless, limited numerical precision in
resolving exponentially small energy scales remains a challenge at
infinite-randomness fixed points. Despite this limitation, our results provide
substantial numerical evidence supporting the Damle--Huse classification of the
random spin-3/2 chain.

\ack{The work was supported by the National Science and Technology Council
(NSTC) in Taiwan under Grants No.~NSTC~114-2119-M-007-013 and NSTC~115-2119-M-007-005.\\
\noindent
{\it AI disclosure statement}\\
\noindent
The code for the numerical domain-wall RG was developed with the assistance of Claude Opus 5.5.
The manuscript was written by the authors. ChatGPT was used to improve language and readability.
The authors reviewed, validated and critically assessed all AI-assisted outputs and take full responsibility for the content of the manuscript.}


\end{document}